\documentclass[12pt]{article}
\usepackage{a4}
\usepackage{amssymb}
\usepackage{cite}
\newcommand{\be}{\begin{equation}}
\newcommand{\ee}{\end{equation}}
\newcommand{\bea}{\begin{eqnarray}}
\newcommand{\eea}{\end{eqnarray}}

\begin{document}
\thispagestyle{empty}
\def\thefootnote{\fnsymbol{footnote}}
\begin{flushright}
{\sf ZMP-HH/09-8}\\
{\sf Hamburger$\;$Beitr\"age$\;$zur$\;$Mathematik$\;$Nr.$\;$ 330}
     \vskip 2em
\end{flushright}
\vskip 2.0em
\begin{center}\Large
Defects and Permutation branes in the Liouville field theory
\end{center}\vskip 1.5em
\begin{center}
Gor Sarkissian
\footnote{\scriptsize
~Email address: \\
$~$\hspace*{2.4em}sarkissian@math.uni-hamburg.de.
}
\end{center}
\begin{center}
Organisationseinheit Mathematik, \ Universit\"at Hamburg\\
Bereich Algebra und Zahlentheorie\\
Bundesstra\ss e 55, \ D\,--\,20\,146\, Hamburg
\end{center}
\vskip 1.5em
\begin{center} March 2009 \end{center}
\vskip 2em
\begin{abstract} \noindent
The defects and permutation branes for the Liouville  field theory are considered.
By exploiting cluster condition, equations satisfied by permutation branes 
and defects reflection amplitudes are obtained.
It is shown that two types of solutions exist, discrete and continuous families.

\end{abstract}

\setcounter{footnote}{0}
\def\thefootnote{\arabic{footnote}}
\newpage

\section{Introduction}
In this paper we address problem of construction permutation branes and topological defects in the 
Liouville field theory.
Topological defects are defined as operators commuting with left and right copies of chiral algebra.
In the last years they were studied extensively in RCFT see, e.g.
\cite{Petkova:2000ip,Petkova:2001ag,Graham:2003nc,Frohlich:2004ef,Frohlich:2006ch,Schweigert:2007wd,Fuchs:2002cm,
Runkel:2007wd}
and free bosonic theory \cite{Bachas:2001vj,Bachas:2007td,Fuchs:2007tx}.
There has been also progress in the Lagrangian description of the defects 
\cite{Fuchs:2007fw,Runkel:2008gr,Sarkissian:2008dq}.

In this paper we turn to construction  of topological defects and closely related 
to them permutation branes in the Liouville field theory.
A discussion of the topological defects in the Liouville theory can be found 
in \cite{Runkel:2007wd}. It was conjectured there that defects in the Liouville theory should be
labelled as FZZ and ZZ branes  \cite{Zamolodchikov:2001ah,Fateev:2000ik} by the primaries  and obey the corresponding fusion rules. 
Our findings here confirm this conjecture.

Main tool used in this paper is generalizations of the Cardy-Lewellen cluster condition
to permutation branes and defects. In the past years Cardy-Lewellen sewing constraint 
proved to be very useful to find branes in non-rational models 
\cite{Zamolodchikov:2001ah,Fateev:2000ik,Giveon:2001uq,Ponsot:2001gt,Lee:2001gh,Teschner:2000md}.
Here we show that for construction of defects in non-rational models it can serve as well.
The paper is organized as follows. 
In section 2 permutation branes in RCFT are reviewed. We collected there necessary
formulae for different annulus partition functions involving permutation branes.
We also elaborate here Cardy-Lewellen cluster conditions for permutation branes.
In section 3 defects in RCFT are reviewed. Here again a special attention to Cardy-Lewellen cluster 
condition for defects is paid.
In section 4 we remind the necessary stuff on Liouville field theory.
In section 5 permutation branes for Liouville theory are presented.
In section 6 defects for Liouville theory are found.

\section{Permutation branes in RCFT}

Let us remind some basic facts on permutation branes in RCFT \cite{Recknagel:2002qq,Gaberdiel:2002jr,Fuchs:2003yk,Sarkissian:2003yw}.
Consider $N$-fold tensor product of a CFT with chiral symmetry algebra $W_L(W_R)$.

On such a product one can consider brane with gluing automorphism given by a cycle 
$(1\ldots N)$, or by other words, satisfying following equations:
\bea\label{perset}
&& W_L^{(r)}(z)=W_R^{(r+1)}(\bar{z})|_{z=\bar{z}},\;\;\; r=1 \ldots  N-1\\ \nonumber
&& W_L^{(N)}(z)=W_R^{(1)}(\bar{z})|_{z=\bar{z}}
\eea
When single copy CFT is a rational CFT
with diagonal partition function 
\be\label{zdiag}
Z=\sum_{i,\bar{i}}Z_{i,\bar{i}}\chi_i(q)\bar{\chi}_{\bar{i}}(\bar{q}),\;\;\; Z_{i,\bar{i}}=\delta_{i,i^*},\;\;\; 
q=\exp(2i\pi\tau)
\ee
where $i^*$ is conjugate representation in the sense $N_{ii^*}^1$=1,
permutation branes were constructed in \cite{Recknagel:2002qq}.
It is shown in \cite{Recknagel:2002qq} that for such a CFT permutation branes are labeled by primaries of single copy and have boundary states:
\begin{equation}
\label{carst}
|a\rangle_{{\cal P}}=\sum_{j}{S_{aj}\over (S_{0j})^{N/2}}|j,j\rangle\rangle_{{\cal P}}
\end{equation}
where $S_{ij}$ is the matrix of the modular transformations of single copy:
\be
\chi_i(\tilde{q})=\sum_j S_{ij}\chi_j(q),\;\;\;\; \tilde{q}=\exp(-2i\pi/\tau)
\ee
and $|j,j\rangle\rangle_{{\cal P}}$ permuted Ishibashi state satisfying (\ref{perset}).
It is known that boundary states should satisfy two criteria: Cardy condition \cite{Cardy:1989ir},
requiring the annulus partition functions to be expressed as sum of some characters
with non-negative integer numbers,
and Cardy-Lewellen cluster condition \cite{Cardy:1991tv,Lewellen:1991tb}.
It is shown in \cite{Recknagel:2002qq} that states (\ref{carst}) indeed satisfy the Cardy condition.
In case of permutation branes check of the Cardy condition involves calculation of two kinds
of annulus partition functions: 

1) partition functions between two permutation branes,

2) partition function between permutation branes and factorized branes, {\it i.e.}
branes which can be written as product of Cardy branes for each constituent.

For further use we write down these partition functions in case of two-fold product $N=2$.
Generalization to generic $N$ is straightforward and corresponding formulae can be found
in \cite{Recknagel:2002qq}.
For two-fold product permutation boundary state (\ref{carst}) satisfies relations:
\bea\label{glrel}
&&L_n^{(1)}-\bar{L}_{-n}^{(2)}=0,\;\;\;\; W_n^{(1)}-(-1)^{s_W}\bar{W}_{-n}^{(2)}=0\\ \nonumber
&&L_n^{(2)}-\bar{L}_{-n}^{(1)}=0,\;\;\;\; W_n^{(2)}-(-1)^{s_W}\bar{W}_{-n}^{(1)}=0
\eea
where $s_W$ is the spin of $W$, and takes  form:
\begin{equation}
\label{per2}
|a\rangle_{{\cal P}}=\sum_{j}{S_{aj}\over S_{0j}}|j,j\rangle\rangle_{{\cal P}}=\sum_{j}{S_{aj}\over S_{0j}}
\sum_{N,M}|j,N\rangle_0\otimes U\overline{|j,N\rangle}_1\otimes|j,M\rangle_1\otimes U\overline{|j,M\rangle}_0
\, .
\end{equation}
where $0$ and $1$ labels first and second copy of the CFT in question, sums over $N$ and $M$ 
run over orthonormal basis of the highest weight representation $R_j$, and operator $U$ in front of right-movers is
chiral CPT operator as usual. 
Using this explicit expression, the Verlinde formula
\be
N_{ij}^k={S_{il}S_{jl}S_{kl}^*\over S_{0l}}
\ee
and expression for the Cardy state:
\be
|i\rangle=\sum_j{S_{ij}\over \sqrt{S_{0j}}}\sum_N |j,N\rangle\otimes U\overline{|j,N\rangle}
\ee
it is easy to compute that 
partition function between two permutation branes labeled by $a_1$ and $a_2$ is :
\be\label{perper}
Z_{a_1,a_2}= \sum_{r,k,l} N^{a_2}_{a_1r}N^r_{kl}\chi_{k}(q)\chi_{l}(q)
\ee

and partition function between permutation brane labeled by $a$ and product of two Cardy states 
labeled by $a_1$ and $a_2$ respectively is ( for details see \cite{Recknagel:2002qq,Sarkissian:2003yw}):
\begin{equation}\label{funcf}
Z_{a,(a_0a_1)}=
\sum_{k,r}N^r_{a_0a_1}N^a_{rk}\chi_{k}(q^{1/2})\, .
\end{equation}

Now we turn to the Cardy-Lewellen cluster condition 
\cite{Cardy:1991tv,Lewellen:1991tb,Pradisi:1996yd,Behrend:1999bn,Runkel:1998pm,Runkel:1999dz}.
Given that cluster condition for permutation branes very little discussed in 
the literature, we will derive it here for general case of the not necessarily diagonal RCFT with
the arbitrary fusion coefficients  $N_{ij}^k$. 
However to keep the things still enough simple we assume that we have no
bulk multiplicities  : $Z_{i\bar{i}}=0,1$. 

Let us as warm-up exercise  to remind cluster condition for usual branes.
Consider a boundary state
\be
|\alpha\rangle=\sum_i B^i_{\alpha}|i\rangle\rangle
\ee
where $i$ runs over primaries, and $|i\rangle\rangle$ are Ishibashi states. 
Recall the relation between coefficients $B^i_{\alpha}$
and one-point functions 
\be
\langle\Phi_{(i\bar{i})}(z,\bar{z})\rangle_{\alpha}={U^i_{\alpha}\delta_{i^*\bar{i}}\over (z-\bar{z})^{2\Delta_i}}
\ee
in the presence of the boundary condition $\alpha$:
\be\label{ub}
U^i_{\alpha}={B^i_{\alpha}\over B^0_{\alpha}}e^{i\pi\Delta_i}
\ee
It is convenient at this point to introduce 
full plane chiral decomposition of physical fields \cite{Moore:1988qv}:
\be\label{decompm}
\Phi_{(i\bar{i})}(z,\bar{z})=\sum_{j,\bar{j},k,\bar{k},a,\bar{a}}C_{(i\bar{i})(j\bar{j})a\bar{a}}^{(k\bar{k})}\left(\phi_{ija}^k(z)\otimes \phi_{\bar{i}\bar{j}\bar{a}}^{\bar{k}}(\bar{z})\right)
\ee
where $\phi_{ija}^k$ are intertwining operators $R_j\rightarrow R_k$, 
and $a=1\ldots N_{ij}^k$.
Consider now two-point function $\langle\Phi_i(z_1,\bar{z}_1)\Phi_j(z_2,\bar{z}_2)\rangle_{\alpha}$ in the presence of boundary in two pictures. In the first picture one applies first bulk OPE
\be
\Phi_{(i\bar{i})}(z_1,\bar{z}_1)\Phi_{(j\bar{j})}(z_2,\bar{z}_2)=\sum_{k,\bar{k}, a, \bar{a}}
{C_{(i\bar{i})(j\bar{j})a\bar{a}}^{(k\bar{k})} \over (z_1-z_2)^{\Delta_i+\Delta_j-\Delta_k}
(\bar{z}_1-\bar{z}_2)^{\Delta_{\bar{i}}+\Delta_{\bar{j}}-\Delta_{\bar{k}}}}
\Phi_{(k\bar{k})}(z_2,\bar{z}_2)+\ldots
\ee
and then evaluates one-point function resulting in:
 \be
\langle\Phi_{(i\bar{i})}(z_1,\bar{z}_1)\Phi_{(j\bar{j})}(z_2,\bar{z}_2)\rangle_{\alpha}=
\sum_{k, a, \bar{a}} C_{(i\bar{i})(j\bar{j})a\bar{a}}^{(k,k^*)}U^k_{\alpha}
{\cal T}_{ka\bar{a}}^{ij\bar{i}\bar{j}}
 \ee
where ${\cal T}_{ka\bar{a}}^{ij\bar{i}\bar{j}}$ are conformal blocks, which  using $\phi_{ija}^k$ intertwining operators
can be expressed as:
\be
{\cal T}_{ka\bar{a}}^{ij\bar{i}\bar{j}}=\langle 0|\phi^1_{ii^*}(z_1)\phi^{i^*}_{jk^*a}(z_2)\phi^{k^*}_{\bar{i}\bar{j}\bar{a}}(\bar{z}_1)
\phi^{\bar{j}}_{\bar{j}1}(\bar{z}_2)|0\rangle
\ee

In the second picture one first applies bulk-boundary OPE \cite{Behrend:1999bn}
\be
\Phi_{(i\bar{i})}(z,\bar{z})=\sum_{m,t,s} {R^{(i\bar{i}), t}_{m,s,(\alpha)}
\over (z-\bar{z})^{\Delta_{i}+\Delta_{\bar{i}}-\Delta_{m}}}
\psi^{\alpha\alpha, s}_m+\ldots
\ee
where $t=1,\ldots N_{i\bar{i}}^m$, and index $s$ counts different boundary fields
and runs $s=1,\ldots n_{\alpha\alpha}^m$, where $n_{\alpha\alpha}^m$ coefficient  
of character $\chi_m$ in the annulus partition function between brane $\alpha$ with itself,
and then evaluates two-point function of boundary fields resulting in
\be
\langle\Phi_{(i\bar{i})}(z_1,\bar{z}_1)\Phi_{(j\bar{j})}(z_2,\bar{z}_2)\rangle_{\alpha}=\sum_{m,t_1,t_2,s_1,s_2}
R^{(i\bar{i}),t_1}_{m,s_1(\alpha)}R^{(j\bar{j}),t_2}_{m^*,s_2(\alpha)}{\cal T}_{mt_1t_2}^{i\bar{i}j\bar{j}}c^{\alpha,s_1,s_2}_m 
\ee
where
\be
\langle\psi^{\alpha\alpha,s_1}_m(x_1)\psi^{\alpha\alpha,s_2}_n(x_2)\rangle={c^{\alpha,s_1,s_2}_m\delta_{mn^*}\over |x_2-x_1|^{2\Delta_m}}
\ee
and
\be
{\cal T}_{mt_1t_2}^{i\bar{i}j\bar{j}}=\langle 0|\phi^1_{ii^*}(z_1)\phi^{i^*}_{\bar{i}m^*t_1}(\bar{z}_1)\phi^{m^*}_{j\bar{j}t_2}(z_2)
\phi^{\bar{j}}_{\bar{j}1}(\bar{z}_2)|0\rangle
\ee
Using braiding relations between chiral blocks
\be\label{matrfm}
{\cal T}_{ka\bar{a}}^{ij\bar{i}\bar{j}}=\sum_m B^{(+)}_{k^*m^*}
 \left[\begin{array}{cc}
j&\bar{i}\\
i^*&\bar{j}\end{array}\right]_{a\bar{a}}^{t_1t_2}
{\cal T}_{mt_1t_2}^{i\bar{i}j\bar{j}}
\ee
one derives:
\be
\sum_{k, a, \bar{a}} C_{(i\bar{i})(j\bar{j})a\bar{a}}^{(k,k^*)}U^k_{\alpha}B^{(+)}_{k^*m^*}
 \left[\begin{array}{cc}
j& \bar{i}\\
i^*&\bar{j}\end{array}\right]_{a\bar{a}}^{t_1t_2}=\sum_{s_1,s_2}R^{(i\bar{i}),t_1}_{m,s_1,(\alpha)}R^{(j\bar{j}),t_2}_{m^*,s_2,(\alpha)}
c^{\alpha,s_1,s_2}_m
\ee

Putting $m=0$ one obtains:

\be
\sum_{k, a, \bar{a}} C_{(ii^*)(jj^*)a\bar{a}}^{(k,k^*)}U^k_{\alpha}B^{(+)}_{k^*0}\left[\begin{array}{cc}
j& i^*\\
i^*& j^*\end{array}\right]_{a\bar{a}}^{11}=U^i_{(\alpha)}U^j_{(\alpha)}
\ee
where we took into account  that $R^{i\bar{i}}_{0(\alpha)}=U^i_{\alpha}\delta_{i^*\bar{i}}$.
We should note that here we used reflection amplitudes  
as they defined in \cite{Behrend:1999bn}. The traditionally used reflection amplitudes 
\cite{Cardy:1991tv,Lewellen:1991tb} differ by phase
\be
U^i_{(\alpha)}=\tilde{U}^i_{(\alpha)}e^{i\pi\Delta_i}
\ee
They have the advantage, that related to boundary states coefficients without phase factor:
\be\label{uclb}
\tilde{U}^i_{(\alpha)}={B^i_{\alpha}\over B^0_{\alpha}}
\ee
Recalling relation between braiding and fusion matrices:
\be\label{brfus}
B_{pq}^{(+)}\left[\begin{array}{cc}
i& j\\
k& l\end{array}\right]_{ab}^{cd}=e^{i\pi(\Delta_k+\Delta_l-\Delta_p-\Delta_q)}
F_{pq}\left[\begin{array}{cc}
i& l\\
k& j\end{array}\right]_{ab}^{cd}
\ee
and symmetry properties of fusion matrix
\be\label{fsym}
F_{pq}\left[\begin{array}{cc}
k& j\\
i& l\end{array}\right]_{ab}^{cd}=F_{p^*q^*}\left[\begin{array}{cc}
l& i^*\\
j^*& k\end{array}\right]_{ab}^{cd}
\ee
we receive that $\tilde{U}^i_{(\alpha)}$ obey the equation:
\be
\sum_{k, a, \bar{a}} C_{(ii^*)(jj^*)a\bar{a}}^{(k,k^*)}\tilde{U}^k_{\alpha}F_{k0}\left[\begin{array}{cc}
i^*& i\\
j& j\end{array}\right]_{a\bar{a}}^{11}=\tilde{U}^i_{(\alpha)}\tilde{U}^j_{(\alpha)}
\ee
Now we apply this procedure to permutation branes. For simplicity we again consider the case of two-fold product.
The primary fields of two-fold product are products of primary fields $\Phi_i^{(1)}\Phi_j^{(2)}$.
The form of the gluing relations (\ref{glrel}) implies that for permutation branes two-point functions have the form:
\be
\langle\Phi_{(i\bar{i})}^{(1)}(z_1)\Phi_{(j\bar{j})}^{(2)}(z_2)\rangle_{\cal P}={U^{i,\bar{i}}_{({\cal P})}
\delta_{i\bar{j}^*}\delta_{\bar{i}j^*}
\over (z_1-\bar{z}_2)^{2\Delta_i}(\bar{z}_1-z_2)^{2\Delta_{\bar{i}}}}
\ee
To receive cluster condition for permutation branes one should consider
four-point functions $\langle\Phi_{(i_1\bar{i}_1)}^{(1)}(z_1)\Phi_{(i_2\bar{i}_2)}^{(2)}(z_2)\Phi_{(j_1\bar{j}_1)}^{(1)}(z_3)
\Phi_{(j_2\bar{j}_2)}^{(2)}(z_4)\rangle_{\cal P}$.
In the first picture one has:
\bea
&&\langle\Phi_{(i_1\bar{i}_1)}^{(1)}(z_1)\Phi_{(i_2\bar{i}_2)}^{(2)}(z_2)\Phi_{(j_1\bar{j}_1)}^{(1)}(z_3)
\Phi_{(j_2\bar{j}_2)}^{(2)}(z_4)\rangle_{\cal P}=\\ \nonumber
&&\sum_{k,\bar{k},a, \bar{a},c,\bar{c}} C_{(i_1\bar{i}_1)(j_1\bar{j}_1)a\bar{a}}^{(k,\bar{k})} C_{(i_2\bar{i}_2)(j_2\bar{j}_2)c\bar{c}}^{(\bar{k}^*,k^*)}
U^{k,\bar{k}}_{({\cal P})} {\cal M}_{k\bar{k}a\bar{a}c\bar{c}}^{i_1i_2j_1j_2\bar{i}_1\bar{i}_2\bar{j}_1\bar{j}_2}
\eea
where ${\cal M}_{k\bar{k}a\bar{a}c\bar{c}}^{i_1i_2j_1j_2\bar{i}_1\bar{i}_2\bar{j}_1\bar{j}_2}$ have the same form as ${\cal T}_k^{ij\bar{i}\bar{j}}$
but with every field being product of two fields for each copy.
Remembering  gluing conditions (\ref{glrel}) we note that actually left fields of
the first copy related only to right fields of the second copy, and right field of the first copy to the left field of the second. Therefore ${\cal M}_{k\bar{k}a\bar{a}c\bar{c}}^{i_1i_2j_1j_2\bar{i}_1\bar{i}_2\bar{j}_1\bar{j}_2}$ factorize and have the form:
\bea
&&{\cal M}_{k\bar{k}a\bar{a}c\bar{c}}^{i_1i_2j_1j_2\bar{i}_1\bar{i}_2\bar{j}_1\bar{j}_2}=\\ \nonumber
&&\langle 0|\phi^1_{i_1i_1^*}(z_1)
\phi^{i_1^*}_{j_1k^*a}(z_3)\phi^{k^*}_{\bar{i}_2\bar{j}_2\bar{c}}(\bar{z}_2)
\phi^{\bar{j}_2}_{\bar{j}_21}(\bar{z}_4)|0\rangle\times\\ \nonumber
&&\langle 0|\phi^1_{i_2i_2^*}(z_2)\phi^{i_2^*}_{j_2\bar{k}c}(z_4)\phi^{\bar{k}}_{\bar{i}_1\bar{j}_1\bar{a}}(\bar{z}_1)
\phi^{\bar{j}_1}_{\bar{j}_11}(\bar{z}_3)|0\rangle={\cal T}_{ka\bar{c}}^{i_1j_1\bar{i}_2\bar{j}_2}{\cal T}_{\bar{k}^*c\bar{a}}^{i_2j_2\bar{i}_1\bar{j}_1}
\eea
Boundary OPE now looks:
\be\label{bope}
\Phi_{(i_1\bar{i}_1)}^{(1)}(z_1)\Phi_{(i_2\bar{i}_2)}^{(2)}(z_2)=\sum_{mn,t_1,t_2,s}
{R^{(i_1\bar{i}_1),(i_2\bar{i}_2),t_1,t_2}_{mn,s}\over
(z_1-\bar{z}_2)^{\Delta_{i_1}+\Delta_{\bar{i}_2}-\Delta_{m}}(\bar{z}_1-z_2)^{\Delta_{\bar{i}_1}+
\Delta_{i_2}-\Delta_{n}}}
\psi_{mn}^s+\ldots
\ee
where $t_1=1\ldots N_{i_1\bar{i}_2}^m$, $t_2=1\ldots N_{\bar{i}_1i_2}^n$, and $s$ counts 
different boundary fields, and its range is given by the corresponding coefficient in the annulus
partition function  of the permutation brane with itself.
Using (\ref{bope}) in the second picture one has:
\bea
&&\langle\Phi_{(i_1\bar{i}_1)}^{(1)}(z_1)\Phi_{(i_2\bar{i}_2)}^{(2)}(z_2)\Phi_{(j_1\bar{j}_1)}^{(1)}(z_3)
\Phi_{(j_2\bar{j}_2)}^{(2)}(z_4)\rangle_{\cal P}=\\ \nonumber
&&\sum_{m,n,t_1,t_2,t_3,t_4,s_1,s_2}R^{(i_1\bar{i}_1),(i_2\bar{i}_2),t_1,t_2}_{mn,s_1}
R^{(j_1\bar{j}_1),(j_2\bar{j}_2),t_3,t_4}_{m^*n^*,s_2}c_{mn}^{s_1,s_2}{\cal M}_{mnt_1t_2t_3t_4}^{i_1i_2\bar{i}_1\bar{i}_2j_1j_2\bar{j}_1\bar{j}_2}
\eea
where
\be
\langle\psi_{mn}^s(x_1)\psi_{pt}^s(x_2)\rangle={c_{mn}^{s_1,s_2}\delta_{mp^*}\delta_{nt^*}\over |x_1-x_2|^{2(\Delta_m+\Delta_n)}}
\ee
and
\bea
&&{\cal M}_{mnt_1t_2t_3t_4}^{i_1i_2\bar{i}_1\bar{i}_2j_1j_2\bar{j}_1\bar{j}_2}=\\ \nonumber
&&\langle 0|\phi^1_{i_1i_1^*}(z_1)\phi^{i_1^*}_{\bar{i}_2m^*t_1}(\bar{z}_2)
\phi^{m^*}_{j_1\bar{j}_2t_3}(z_3)
\phi^{\bar{j}_2}_{\bar{j}_21}(\bar{z}_4)|0\rangle\times\\ \nonumber
&&\langle 0|\phi^1_{i_2i^*_2}(z_2)\phi^{i_2^*}_{\bar{i}_1n^*t_2}(\bar{z}_1)\phi^{n^*}_{j_2\bar{j}_1t_4}(z_4)
\phi^{\bar{j}_1}_{\bar{j}_11}(\bar{z}_3)|0\rangle={\cal T}_{mt_1t_3}^{i_1\bar{i}_2j_1\bar{j}_2}
{\cal T}_{nt_2t_4}^{i_2\bar{i}_1j_2\bar{j}_1}
\eea
Using (\ref{matrfm}) we end up with:
\bea
\sum_{k,\bar{k},a, \bar{a},c,\bar{c}} C_{(i_1\bar{i}_1)(j_1\bar{j}_1)a\bar{a}}^{(k,\bar{k})} 
C_{(i_2\bar{i}_2)(j_2\bar{j}_2)c\bar{c}}^{(\bar{k}^*,k^*)}B^{(+)}_{k^*m^*}
 \left[\begin{array}{cc}
j_1&\bar{i}_2\\
i_1^*&\bar{j}_2\end{array}\right]_{a\bar{c}}^{t_1t_3}B^{(+)}_{\bar{k}n^*}
 \left[\begin{array}{cc}
j_2&\bar{i}_1\\
i_2^*&\bar{j}_1\end{array}\right]_{c\bar{a}}^{t_2t_4}
U^{k,\bar{k}}_{({\cal P})}=\\ \nonumber
\sum_{s_1,s_2}R^{(i_1\bar{i}_1),(i_2\bar{i}_2),t_1,t_2}_{mn,s_1}
R^{(j_1\bar{j}_1),(j_2\bar{j}_2),t_3,t_4}_{m^*n^*,s_2}c_{mn}^{s_1,s_2}
\eea

 Putting $m=n=0$, and taking into account that
 \be
 R^{(i_1\bar{i}_1),(i_2\bar{i}_2),t_1,t_2}_{00,s}=U^{i_1,\bar{i}_1}_{({\cal P})}\delta_{i_1^*\bar{i_2}}
\delta_{i_2\bar{i_1}^*}
\ee
 one obtains:
 
 \bea\label{clperr}
\sum_{k,\bar{k},a, \bar{a},c,\bar{c}} C_{(i_1\bar{i}_1)(j_1\bar{j}_1)a\bar{a}}^{(k,\bar{k})} 
C_{(\bar{i}_1^*i_1^*)(\bar{j}_1^*j_1^*)c\bar{c}}^{(\bar{k}^*,k^*)}B^{(+)}_{k^*0}
 \left[\begin{array}{cc}
j_1&i_1^*\\
i_1^*&j_1^*\end{array}\right]_{a\bar{c}}^{11}B^{(+)}_{\bar{k}0}
 \left[\begin{array}{cc}
\bar{j}_1^*&\bar{i}_1\\
\bar{i}_1&\bar{j}_1\end{array}\right]_{c\bar{a}}^{11}
U^{k,\bar{k}}_{({\cal P})}=\\ \nonumber
U^{i_1,\bar{i}_1}_{({\cal P})}U^{j_1,\bar{j}_1}_{({\cal P})}
\eea
 
 Again defining new amplitudes
 \be
 \tilde{U}^{i_1,\bar{i}_1}_{({\cal P})}=U^{i_1,\bar{i}_1}_{({\cal P})}e^{i\pi(\Delta_i+\Delta_{\bar{i}})}
 \ee
 and using (\ref{brfus}) and (\ref{fsym})
 we derive:
 \bea\label{clper}
\sum_{k,\bar{k},a, \bar{a},c,\bar{c}} C_{(i_1\bar{i}_1)(j_1\bar{j}_1)a\bar{a}}^{(k,\bar{k})} 
C_{(\bar{i}_1^*i_1^*)(\bar{j}_1^*j_1^*)c\bar{c}}^{(\bar{k}^*,k^*)}F_{k0}
 \left[\begin{array}{cc}
i_1^*&i_1\\
j_1^*&j_1\end{array}\right]_{a\bar{c}}^{11}F_{\bar{k}^*0}
 \left[\begin{array}{cc}
\bar{i}_1&\bar{i}_1^*\\
\bar{j}_1^*&\bar{j}_1^*\end{array}\right]_{c\bar{a}}^{11}
\tilde{U}^{k,\bar{k}}_{({\cal P})}=\\ \nonumber
\tilde{U}^{i_1,\bar{i}_1}_{({\cal P})}\tilde{U}^{j_1,\bar{j}_1}_{({\cal P})}
\eea
 For diagonal model $i_1=\bar{i}_1^*$, $j_1=\bar{j}_1^*$, $\bar{k}=k^*$ without multiplicities
 $N_{ij}^k=1$, (\ref{clper}) simplifies to

\be\label{clustper}
\sum_k (C_{ij}^k)^2 \tilde{U}^k_{({\cal P})}\left(F_{k0}\left[\begin{array}{cc}
i^*&i\\
j&j\end{array}\right]\right)^2=\tilde{U}^i_{({\cal P})}\tilde{U}^j_{({\cal P})}
\ee
were we denoted $C_{ij}^k\equiv C_{(ii^*)(jj^*)}^{(kk^*)}$ and $\tilde{U}^i_{({\cal P})}\equiv \tilde{U}^{i,i^*}_{({\cal P})}$.

Note that for diagonal models permutation branes reflection amplitudes depend  
only on single copy primaries. 

For this case permutation branes cluster condition was discussed in \cite{Recknagel:2002qq}.

It is straightforward to generalize (\ref{clper}) to general $N$-fold product.
Here we only write the corresponding  formula for diagonal models (\ref{zdiag}) without
multiplicities:

\be\label{clustpern}
\sum_k (C_{ij}^k)^N \tilde{U}^k_{({\cal P})}\left(F_{k0}\left[\begin{array}{cc}
i^*&i\\
j&j\end{array}\right]\right)^N=\tilde{U}^i_{({\cal P})}\tilde{U}^j_{({\cal P})}
\ee

It is shown in \cite{Recknagel:2002qq} that (\ref{carst}) satisfies (\ref{clustpern}).

\section{Topological defects in RCFT}

Recall basic facts on topological defects in RCFT 
 \cite{Petkova:2000ip,Petkova:2001ag,Graham:2003nc,Fuchs:2002cm}.
The construction of defects lines is analogous to that of boundary condition. Following 
\cite{Petkova:2000ip} we define defect lines as operators $X$, satisfying relations:
\be\label{lint}
[L_n,X]=[\bar{L}_n,X]=0
\ee
\be\label{wint}
[W_n,X]=[\bar{W}_n,X]=0
\ee

As in the case of the boundary conditions, there are also consistency conditions, analogous
to the Cardy and Cardy-Lewellen constraints,
which must be satisfied by the operator $X$. For simplicity we shall write all the formulae
for diagonal models (\ref{zdiag}).
To formulate these conditions, one first note that as consequence of (\ref{lint}) and (\ref{wint})
$X$ is a sum of projectors
\be\label{xpd}
X=\sum_{i,\bar{i}}{\cal D}^{(i,\bar{i})}P^{(i,\bar{i})}
\ee
where
\be
P^{(i,\bar{i})}=\sum_{N,\bar{N}}(|i,N\rangle\otimes |\bar{i},\bar{N}\rangle)
(\langle i,N|\otimes \langle\bar{i},\bar{N}|)
\ee
An analogue of the Cardy condition for defects requires that partition function with insertion of a pair defects
after modular transformation can be expressed as sum of characters with non-negative integers.
It is found in  \cite{Petkova:2000ip} that for diagonal models one can solve this condition 
taking for each primary $a$
\be\label{top}
{\cal D}^{(i,\bar{i})}_a={S_{ai}\over S_{0i}}
\ee
for which one has:
\be\label{deffu}
Z_{ab}={\rm Tr}\left(X^{\dagger}_aX_b\tilde{q}^{L_0-{c\over 24}}\tilde{\bar{q}}^{\bar{L}_0-{c\over 24}}\right)=
\sum_{k,i\bar{i}}N^a_{bk}N^k_{i\bar{i}}\chi_{i}(q)\chi_{\bar{i}}(\bar{q})
\ee

Topological defects can act on boundary states producing new boundary states.
The action of defects (\ref{top}) on Cardy states is easily obtained using the Verlinde formula:
\be\label{dbs}
X_a|b\rangle=\sum_d N^d_{ab}|d\rangle
\ee
Topological defects can be fused. For defects (\ref{top}) again using the Verlinde formula one derives:
\be\label{dds}
X_aX_b=\sum_c N^c_{ab}X_c
\ee
Now we turn to the cluster condition for defects \cite{Petkova:2001ag}.
Here we should consider two-point functions 
\be
\langle\Phi_{i^*}(z_1,\bar{z}_1)X\Phi_{i}(z_2,\bar{z}_2)\rangle={D^{(i,\bar{i})}\over 
(z_1-z_2)^{2\Delta_i}(\bar{z}_1-\bar{z}_2)^{2\Delta_{\bar{i}}}}
\ee
\be
D^i={{\cal D}^{(i,\bar{i})}\over {\cal D}^0}
\ee

Using (\ref{decompm}) one can write for the following four-point function with the defects insertion 
 in the first picture:
\bea\label{frprd}
&&\langle\Phi_{j^*}(z_1,\bar{z}_1)\Phi_{i^*}(z_2,\bar{z}_2)X\Phi_{i}(z_3,\bar{z}_3)\Phi_{j}(z_4,\bar{z}_4)X^{\dagger}\rangle=\\ \nonumber
&&\sum_k C_{j^*j}^1C_{ij,a\bar{a}}^{k}C_{i^*k,c\bar{c}}^{j}
D^{(k,\bar{k})}{\cal F}_{kac}^{j^*i^*ij}{\cal F}_{\bar{k}\bar{a}\bar{c}}^{\bar{j}^*\bar{i}^*\bar{i}\bar{j}}
\eea
where
\be\label{bl1m}
{\cal F}_{kac}^{j^*i^*ij}=\langle 0|\phi^1_{j^*j}(z_1)\phi^j_{i^*kc}(z_2)\phi^k_{ija}(z_3)\phi^j_{j1}(z_4)|0\rangle
\ee
Here we denoted $C_{ij}^k\equiv C_{(ii^*)(jj^*)}^{(kk^*)}$ as before.

Using relations:
\be
C_{i^*k,c\bar{c}}^{j}=
C_{ki^*,c\bar{c}}^{j}
\ee

and
\be
C_{ki^*,c\bar{c}}^{j}C_{j^*j}^1=
C_{i^*j^*,c\bar{c}}^{k^*}C_{k^*k}^1
\ee
we can write for the second line of (\ref{frprd})
\be
\sum_k C_{ij,a\bar{a}}^{k}C_{i^*j^*,c\bar{c}}^{k^*}C_{k^*k}^1D^{(k,\bar{k})}{\cal F}_{kac}^{j^*i^*ij}{\cal F}_{\bar{k}\bar{a}\bar{c}}^{\bar{j}^*\bar{i}^*\bar{i}\bar{j}}
\ee
In the second picture one has:
\bea\label{secprd}
\langle\Phi_{i^*}(z_2,\bar{z}_2)X\Phi_{i}(z_3,\bar{z}_3)\Phi_{j}(z_4,\bar{z}_4)X^{\dagger}\Phi_{j^*}(z_1,\bar{z}_1)\rangle=\\ \nonumber
C_{i^*i}^1C_{j^*j}^1D^{(i,\bar{i})}D^{(j,\bar{j})}{\cal F}_0^{i^*ijj^*}{\cal F}_0^{\bar{i}^*\bar{i}\bar{j}\bar{j}^*}+\ldots
\eea
where

\be\label{bl2m}
{\cal F}_{pmn}^{i^*ijj^*}=\langle 0|\phi^1_{i^*i}(z_2)\phi^i_{ipm}(z_3)\phi^p_{jj^*n}(z_4)\phi^{j^*}_{j^*1}(z_1)|0\rangle
\ee
To relate (\ref{frprd}) with (\ref{secprd}) one should use braiding relations for chiral blocks to move $j^*$ to the very right.
Using (\ref{brfus}) and the following property of the braiding matrix 
\be
B_{ij}^{(+)}\left[\begin{array}{cc}
i^*& j^*\\
0& k\end{array}\right]_{1a}^{1a}=(\pm)e^{i\pi(\Delta_k-\Delta_i-\Delta_j)}
\ee
one obtains product of fusion matrices : 
\be
{\cal F}_{kac}^{j^*i^*ij}{\cal F}_{\bar{k}\bar{a}\bar{c}}^{\bar{j}^*\bar{i}^*\bar{i}\bar{j}}=
 F_{k0}\left[\begin{array}{cc}
j^*&j\\
i&i\end{array}\right]_{ac}^{11} F_{\bar{k}0}\left[\begin{array}{cc}
\bar{j}^*&\bar{j}\\
\bar{i}&\bar{i}\end{array}\right]_{\bar{a}\bar{c}}^{11}{\cal F}_{0}^{i^*ijj^*}{\cal F}_0^{\bar{i}^*\bar{i}\bar{j}\bar{j}^*}+\ldots
\ee
Collecting all we obtain
\bea\label{defalg}
\sum_k (C_{k^*k}^1D^{(k\bar{k})})C_{ij,a\bar{a}}^{k}
C_{i^*j^*,c\bar{c}}^{k^*}
 F_{k0}\left[\begin{array}{cc}
j^*&j\\
i&i\end{array}\right]_{ac}^{11}F_{\bar{k}0}\left[\begin{array}{cc}
\bar{j}^*&\bar{j}\\
\bar{i}&\bar{i}\end{array}\right]_{\bar{a}\bar{c}}^{11}=\\ \nonumber
(C_{i^*i}^1D^{(i\bar{i})})(C_{(j^*j}^1D^{(j\bar{j})})
\eea
Comparing formulae (\ref{glrel}) and (\ref{lint}), (\ref{per2}) and (\ref{xpd}), (\ref{top}),
(\ref{clper}) and (\ref{defalg}) one reveals deep connection between permutation branes on two-fold product
form one side, and defects on other side, known as folding trick 
\cite{Bachas:2001vj,Bachas:2007td,Oshikawa:1996dj,Wong:1994pa}.
We see that mentioned relations for permutation branes become corresponding relations for defect
after performing two-steps operation (folding) on the second copy of the CFT in question: left-right exchange 
and then hermitian conjugation, turning boundary state to operator.
Comparison of (\ref{clper}) and (\ref{defalg}) shows that the hermitian conjugation requires
inclusion  of the  two-point functions $C_{i^*i}^1$.

\section{Liouville theory}
Let us review basic facts on the Liouville field theory (see e.g. \cite{Teschner:2001rv}).
Liouville field theory is defined on a two-dimensional surface with metric $g_{ab}$ by the local Lagrangian
density
\be
{\cal L}={1\over 4\pi}g_{ab}\partial_a\varphi\partial_b \varphi+\mu e^{2b\varphi}+{Q\over 4\pi}R\varphi
\ee
where $R$ is associated curvature. This theory is conformal invariant if the coupling constant $b$
is related with the background charge $Q$ as
\be
Q=b+{1\over b}
\ee

The symmetry algebra of this conformal field theory is the Virasoro algebra
\be
[L_m,L_n]=(m-n)L_{m+n}+{c_L\over 12}(n^3-n)\delta_{n,-m}
\ee
with central charge 
\be
c_L=1+6Q^2
\ee

Primary fields $V_{\alpha}$ in this theory, which are associated with exponential fields
$e^{2\alpha \varphi}$, have conformal dimensions
\be
\Delta_{\alpha}=\alpha(Q-\alpha)
\ee

The fields $V_{\alpha}$ and $V_{Q-\alpha}$ have the same conformal dimensions and represent
the same primary field, i.e. they are proportional to each other:
\be
V_{\alpha}=S(\alpha)V_{Q-\alpha}
\ee
with the function
\be\label{reflal}
S(\alpha)={\left(\pi\mu\gamma(b^2)\right)^{b^{-1}(Q-2\alpha)}\over b^2}{\Gamma(1-b(Q-2\alpha))\Gamma(-b^{-1}(Q-2\alpha))
\over \Gamma(b(Q-2\alpha))\Gamma(1+b^{-1}(Q-2\alpha))}
\ee

The spectrum of the Liouville theory is believed \cite{Curtright:1982gt,Braaten:1982yn,Braaten:1983np}
to be of the following form
\be\label{lsdi}
{\cal H}=\int_0^{\infty} dp \;R_{{Q\over 2}+iP}\otimes R_{{Q\over 2}+iP}
\ee
where $R_{\alpha}$ is the highest weight representation with respect to Virasoro alegbra.
Characters of the representations $R_{{Q\over 2}+iP}$ are
\be\label{charl}
\chi_{P}(\tau)={q^{P^2}\over \eta(\tau)}
\ee
where 
\be
\eta(\tau)=q^{1/24}\prod_{n=1}^{\infty}(1-q^n)
\ee
Modular transformation of (\ref{charl}) is well-known:
\be\label{motrp}
\chi_{P}(-{1\over \tau})=\sqrt{2}\int\chi_{P'}(\tau)e^{4i\pi PP'}dP'
\ee
Degenerate representations appear at $\alpha_{m,n}={1-m\over 2b}+{1-n\over 2}b$ and have
conformal dimensions \cite{Kac:1990gs}
\be
\Delta_{m,n}=Q^2/4-(m/b+nb)^2/4
\ee
where $m,n$ are positive integers. At general $b$ there is only one null-vector at the level $mn$.
Hence the degenerate character reads:
\be\label{charmn}
\chi_{m,n}(\tau)={q^{-(m/b+nb)^2}-q^{-(m/b-nb)^2}\over \eta(\tau)}
\ee

Modular transformation of (\ref{charmn}) is worked out in \cite{Zamolodchikov:2001ah}
\be\label{motrdi}
\chi_{m,n}(-{1\over \tau})=2\sqrt{2}\int\chi_{P}(\tau)\sinh(2\pi mP/b)\sinh(2\pi nbP)dP
\ee

For future use we write here the reflection function 
for $\alpha={Q\over 2}+iP$, denoting it briefly as $S(P)$:

\be\label{reflp}
S(P)=-[\pi\mu\gamma(b^2)]^{-i2P/b}{\Gamma(1+2ibP)\Gamma(1+{2iP\over b})\over \Gamma(1-2ibP)\Gamma(1-{2iP\over b})}
\ee
Two-point functions of Liouvulle theory are given by reflection function 
(\ref{reflal}):
\be\label{twopf}
\langle V_{\alpha}(z_1,\bar{z}_1)V_{\alpha}(z_2,\bar{z}_2)\rangle={S(\alpha)\over (z_1-z_2)^{2\Delta_{\alpha}}
(\bar{z}_1-\bar{z}_2)^{2\Delta_{\alpha}}}
\ee
Three-point functions of Liouville theory $C(\alpha_1,\alpha_2,\alpha_3)$ are computed 
in \cite{Dorn:1994xn,Zamolodchikov:1995aa}, were so called DOZZ formula for them was suggested. 
We don't need in this paper the full DOZZ formula.
But we do need $C(\alpha_1,\alpha_2,\alpha_3)$ for the values of $\alpha_i$ satisfying relation
\be
\alpha_1+\alpha_2+\alpha_3=Q-nb
\ee
For this case three-point functions are given by the screening integrals computed in \cite{Dotsenko:1984ad}
\be\label{df}
I_n(\alpha_1,\alpha_2,\alpha_3)=\left(b^4\gamma(b^2)\pi\mu\right)^n
{\prod_{j=1}^n\gamma(-jb^2)\over \prod_{k=0}^{n-1}[\gamma(2\alpha_1 b+kb^2)\gamma(2\alpha_2 b+kb^2)
\gamma(2\alpha_3 b+kb^2)]}
\ee

where $\gamma(x)={\Gamma(x)\over \Gamma(1-x)}$.

Structure constants $C^{\alpha_3}_{\alpha_1,\alpha_2}$ are related to three-point functions as
\be\label{struc}
C^{\alpha_3}_{\alpha_1,\alpha_2}=C(\alpha_1,\alpha_2,Q-\alpha_3)
\ee

\section{Permutation branes in Liouville theory}
In this section we turn to construction of permutation branes on $N$-fold product of the Liouville field
theories. As explained in section 1 they satisfy following gluing conditions:
\bea\label{virgl}
&&L_{n}^{(r)}-\bar{L}_{-n}^{(r)}=0,\;\;\; r=1\ldots N-1,\\ \nonumber
&&L_{n}^{(N)}-\bar{L}_{-n}^{(1)}=0.
\eea
Remembering that Liouville field theory is diagonal theory (\ref{lsdi})
we conclude that reflection amplitudes as well as Ishibashi states depend
on single copy primaries $P$.
To compute reflection amplitudes $U_{{\cal P}}^{(N)}(\alpha={Q\over 2}+iP)$
for permutation branes on $N$-fold product
of Liouville fields
\be\label{onepo}
\langle V_{{Q\over 2}+iP}^{(1)}(z_1,\bar{z}_1)\cdots V_{{Q\over 2}+iP}^{(N)}(z_N,\bar{z}_N)\rangle_{{\cal P}}=
{U_{{\cal P}}^{(N)}(\alpha)\over \prod_1^{N}(z_r-\bar{z}_{r+1})^{(Q^2/2+2P^2)}}
\ee
where $z_{N+1}\equiv z_1$,
we will use the same trick
as in \cite{Teschner:1995yf,Fateev:2000ik,Zamolodchikov:2001ah} and apply sewing constraints to
$2N$-point function
\be
\langle V_{-b/2}^{(1)}(z_1,\bar{z}_1)\cdots V_{-b/2}^{(N)}(z_N,\bar{z}_N)
 V_{{Q\over 2}+iP}^{(1)}(z_{N+1},\bar{z}_{N+1})\cdots V_{{Q\over 2}+iP}^{(N)}(z_{2N},\bar{z}_{2N})\rangle_{{\cal P}}
 \ee
with degenerate representation $-b/2$.
Recalling fusion rule with degenerate field

\be\label{fusv}
V_{-b/2}V_{\alpha}\sim C_{-b/2,\alpha}^{\alpha-b/2}V_{\alpha-b/2}+C_{-b/2,\alpha}^{\alpha+b/2}V_{\alpha+b/2}
\ee
and that Liouville theory is diagonal theory with self-conjugate primaries 
we can apply to this situation 
equation (\ref{clustpern}) with $i=-b/2$, $j=\alpha={Q\over 2}+iP$, $k=\alpha\pm b/2$:
\bea\label{baseq}
&&U_{{\cal P}}^{(N)}(\alpha)U_{{\cal P}}^{(N)}(-b/2)=\\ \nonumber
&&\left(C_{-b/2,\alpha}^{\alpha-b/2}F_{\alpha-b/2,0}\left[\begin{array}{cc}
-b/2&-b/2\\
\alpha &\alpha \end{array}\right]\right)^N U_{{\cal P}}^{(N)}(\alpha-b/2)\\ \nonumber
&&+\left(C_{-b/2,\alpha}^{\alpha+b/2}F_{\alpha+b/2,0}\left[\begin{array}{cc}
-b/2&-b/2\\
\alpha&\alpha\end{array}\right]\right)^N U_{{\cal P}}^{(N)}(\alpha+b/2)
\eea
The necessary three-point functions can be computed using (\ref{df}) and (\ref{struc})
\be\label{c1b}
C_{-b/2,\alpha}^{\alpha-b/2}=C(-b/2,\alpha,Q-\alpha+b/2)=1
\ee
\be\label{c2b}
C_{-b/2,\alpha}^{\alpha+b/2}=C(\alpha, -b/2,Q-\alpha-b/2)=
b^4\pi\mu \gamma(b^2){\Gamma(2\alpha b-b^2-1)\Gamma(1-2\alpha b)\over \Gamma(2+b^2-2\alpha b)\Gamma(2\alpha b)}
\ee
The necessary elements of the fusion  matrix are computed in \cite{Teschner:1995yf,Fateev:2000ik,Zamolodchikov:2001ah}
using explicit expression of the conformal blocks through  hypergeometric functions.
We will write down here final results:
\be\label{fb1}
F_{\alpha-b/2,0}\left[\begin{array}{cc}
-b/2&-b/2\\
\alpha &\alpha \end{array}\right]={\Gamma(2\alpha b-b^2)\Gamma(-1-2b^2)\over 
\Gamma(2\alpha b-2b^2-1)\Gamma(-b^2)}
\ee
\be\label{fb2}
F_{\alpha+b/2,0}\left[\begin{array}{cc}
-b/2&-b/2\\
\alpha &\alpha \end{array}\right]={\Gamma(2+b^2-2\alpha b)\Gamma(-1-2b^2)\over 
\Gamma(1-2\alpha b)\Gamma(-b^2)}
\ee

At this point we can continue in two different ways.
It is shown in \cite{Fateev:2000ik,Zamolodchikov:2001ah} that Liouville theory possesses two kinds
of boundary states, discrete and continuous families.
For permutation branes and defects one expects the same picture.
To discover continuous family one treats $U_{{\cal P}}^{(N)}(-b/2)$ as 
a constant $A$ depending on boundary condition. Doing this and putting (\ref{c1b}), (\ref{c2b}),
(\ref{fb1}) and (\ref{fb2}) in (\ref{baseq}) one receives the following linear equation:
\bea\label{lineq}
&& AU_{{\cal P}}^{(N)}(\alpha)=\left({\Gamma(-1-2b^2)\Gamma(2\alpha b-b^2)\over
\Gamma(-b^2)\Gamma(2\alpha b-2b^2-1)}\right)^N U_{{\cal P}}^{(N)}(\alpha-b/2)+\\ \nonumber
&&\left({\pi \mu \gamma(b^2)b^4\Gamma(-1-2b^2)\Gamma(2\alpha b-b^2-1)\over
\Gamma(-b^2)\Gamma(2\alpha b)}\right)^N U_{{\cal P}}^{(N)}(\alpha+b/2)
\eea

Using the identity 
\be\label{gam}
\Gamma(1+z)=z\Gamma(z)
\ee
 it is easy to show that (\ref{lineq})
can be solved by:

\be
U_{{\cal P}\, s}^{(N)}(\alpha)=2^{1/2}\left[{1\over 2^{3/4}\pi b}(\pi\mu\gamma(b^2))^{(Q-2\alpha)/2b}\Gamma(1-b(Q-2\alpha))\Gamma(-b^{-1}(Q-2\alpha))\right]^N
\cosh(2\pi s(2\alpha-Q))
\ee

where
\be
2\cosh 2\pi bs={A\over b^{2N}}\left({\Gamma(-b^2)\over \Gamma(-1-2b^2)}\right)^N{1\over (\pi\mu\gamma(b^2))^{N/2}}
\ee

Putting $\alpha={Q\over 2}+iP$ we get
\be
\label{permtwo}
U_{{\cal P}\, s}^{(N)}(P)=2^{1/2}\left({[\pi\mu\gamma(b^2)]^{-iP/b}\Gamma(1+2ibP)\Gamma(1+{2iP\over b})\over 2^{3/4}(2i\pi P)}\right)^N
\cos(4P\pi s)
\ee

Let us make the following comments on (\ref{permtwo}).
\begin{enumerate}
\item Putting $N=1$ we surely recover FZZ branes\footnote{To compare with  \cite{Fateev:2000ik}
we changed here slightly normalization, and redefined 
parameter $s$ there as $2s$ here.}
 \cite{Fateev:2000ik}:
\be
U_{{\cal P}\, s}^{(1)}(P)\equiv U_{ s}^{(FZZ)}(P)={2^{-1/4}[\pi\mu\gamma(b^2)]^{-iP/b}\Gamma(1+2ibP)\Gamma(1+{2iP\over b})\over 2i\pi P}
\cos(4P\pi s)
\ee
\item
It is very interesting to note that (\ref{permtwo}) has similar structure as 
corresponding solution (\ref{carst}) in the case of rational CFT, in the sense
that both have the form $S_{aj}(f(j))^N$, where $S_{aj}$ is the matrix of the modular transformation
of the single copy, and $f(j)$ is the function which appears in the expression for single copy boundary states. 

\item  From the expression (\ref{onepo}) one concludes that $U_{{\cal P}}^{(N)}(P)$
should  satisfy 
\be\label{refun}
U_{{\cal P}\, s}^{(N)}(P)=(S(P))^NU_{{\cal P}\, s}^{(N)}(-P)
\ee
Solution (\ref{permtwo}) obviously satisfies (\ref{refun}).

\end{enumerate}

To obtain discrete family we will treat $U_{{\cal P}}^{(N)}(-b/2)$ as it stands, and again
substituting in (\ref{baseq}) values of structure constants and elements of fusion matrix
(\ref{c1b}), (\ref{c2b}),
(\ref{fb1}) and (\ref{fb2}), we derive the following non-linear equation:
\bea\label{nonlin}
&&U_{{\cal P}}^{(N)}(\alpha)U_{{\cal P}}^{(N)}(-b/2)=\left({\Gamma(-1-2b^2)\Gamma(2\alpha b-b^2)\over
\Gamma(-b^2)\Gamma(2\alpha b-2b^2-1)}\right)^N U_{{\cal P}}^{(N)}(\alpha-b/2)+\\ \nonumber
&&\left({\pi \mu b^4\gamma(b^2)\Gamma(-1-2b^2)\Gamma(2\alpha b-b^2-1)\over
\Gamma(-b^2)\Gamma(2\alpha b)}\right)^N U_{{\cal P}}^{(N)}(\alpha+b/2)
\eea
Equation (\ref{nonlin}) admits  the following two-parameters solution:
\be
U_{{\cal P}\,m,n}^{(N)}(\alpha)=\left(
{[\pi\mu\gamma(b^2)]^{-\alpha/b}\Gamma(1-b(Q-2\alpha))\Gamma(-b^{-1}(Q-2\alpha))\over \Gamma(1-bQ)\Gamma(-b^{-1}Q)}\right)^N f_{m,n}(\alpha)
\ee

where 
\be
f_{m,n}(\alpha)={\sin(\pi m b^{-1}(2\alpha- Q))\sin(\pi nb (2\alpha-Q))\over \sin(\pi mb^{-1} Q)\sin(\pi nb Q)}
\ee
and satisfies equation
\be
f_{m,n}(\alpha)f_{m,n}(-b/2)=f_{m,n}(\alpha-b/2)+f_{m,n}(\alpha+b/2)
\ee
Putting $\alpha={Q\over 2}+iP$ we get

\be\label{solmn}
U_{{\cal P}\,m,n}^{(N)}(P)=\left(
{[\pi\mu\gamma(b^2)]^{-Q/2b}[\pi\mu\gamma(b^2)]^{-iP/b}\Gamma(1+2iPb)\Gamma(1+2iP/b)Q\over \Gamma(1-bQ)\Gamma(1-b^{-1}Q)(-2iP)}\right)^N f_{m,n}(P)
\ee
where 
\be
f_{m,n}(P)={\sinh(2\pi mP b^{-1})\sinh(2\pi nbP)\over \sin(\pi mb^{-1} Q)\sin(\pi nb Q)}
\ee

To construct boundary states one should solve additionally the equation (\ref{uclb}).
The solution is easily seen to be
\be\label{defps}
\Psi_{{\cal P}\,m,n}^{(N)}(P)=2^{3/2}\left(
{[\pi\mu\gamma(b^2)]^{-iP/b}\Gamma(1+2iPb)\Gamma(1+2iP/b)\over 2^{3/4}(2i\pi P)}\right)^N \sinh(2\pi mP b^{-1})\sinh(2\pi nbP)
\ee

\be
U_{{\cal P}\,m,n}^{(N)}(P)={\Psi_{{\cal P}\,m,n}^{(N)}(P)\over \Psi_{{\cal P}\,m,n}^{(N)}(i{Q\over 2})}
\ee

For solution (\ref{defps}) we can make similar comments as for solution (\ref{permtwo}).
For $N=1$ we recover ZZ branes:
\bea
&&\Psi_{{\cal P}\,m,n}^{(1)}(P)\equiv \Psi_{m,n}^{(ZZ)}(P)=\\ \nonumber
&&{2^{3/4}[\pi\mu\gamma(b^2)]^{-iP/b}\Gamma(1+2iPb)\Gamma(1+2iP/b)\over 2i\pi P}\sinh(2\pi mP b^{-1})\sinh(2\pi nbP)
\eea
The solution (\ref{defps}) has the same structure as (\ref{carst}) in the same sense as before ,
and satisfies the reflection constraint (\ref{refun}).

Having reflection amplitudes  (\ref{permtwo}) and  (\ref{defps}) one can write boundary states
\be\label{sN}
|s\rangle^{(N)}_{{\cal P}}=\int U_{{\cal P}\, s}^{(N)}(P)|P\rangle\rangle^{(N)}_{{\cal P}}dP
\ee
\be\label{mnN}
|m,n\rangle^{(N)}_{{\cal P}}=\int\Psi_{{\cal P}\,m,n}^{(N)}(P) |P\rangle\rangle^{(N)}_{{\cal P}}dP
\ee
where $|P\rangle\rangle^{(N)}_{{\cal P}}$ are Ishibashi states satisfying (\ref{virgl}).
For $N=1$ we identify  
\be
|s\rangle^{(1)}_{{\cal P}}\equiv |s\rangle^{(FZZ)}=\int U_{ s}^{(FZZ)}(P)|P\rangle\rangle dP
\ee

\be
|m,n\rangle^{(1)}_{{\cal P}}\equiv |m,n\rangle^{(ZZ)}=\int \Psi_{m,n}^{(ZZ)}(P)|P\rangle\rangle dP
\ee
where $|P\rangle\rangle$ are the Ishibashi states satisfying  $L_n+\bar{L}_{-n}=0$.

Let us test the solutions (\ref{sN}) and (\ref{mnN})  computing the annulus partition function between permutation branes and products of ZZ branes.
 For simplicity we restrict ourselves to the case
of permutation branes on two-fold product $N=2$.
The partition function between permutation brane labelled by $s$
and product of two ZZ branes labelled by $(m_1,n_1)$ and $(m_2,n_2)$ respectively is
\bea\label{psmn}
&&Z_{s,(m_1,n_1),(m_2,n_2)}=\int U_{{\cal P}\, s}^{(2)}(-P)\Psi_{m_1,n_1}^{ZZ}(P)\Psi_{m_2,n_2}^{(ZZ)}(P)(\chi_P(\tilde{q}))^2dP=\\ \nonumber
&&\int_P {2^{1/2}\cos(4P\pi s)\sinh(2\pi m_1bP)\sinh(2\pi n_1P/b)\sinh(2\pi m_2bP)\sinh(2\pi n_2P/b)
\over (\sinh(2bP)\sinh (2P/b))^2}\chi_P(\tilde{q}^2)dP
\eea
To obtain (\ref{psmn}) we used the $\Gamma$-function identity
\be\label{giden}
\Gamma(1+ix)\Gamma(1-ix)={\pi x\over \sinh(\pi x)}
\ee
 
and the following property of the permutation Ishibashi states 
\be\label{ishid}
\langle\langle Q_1|\langle\langle Q_2 |(\tilde{q}^{1/2})^H|P\rangle\rangle^{(2)}_{{\cal P}}=
\chi_P(\tilde{q}^{2})\delta(P-Q_1)\delta(P-Q_2)
\ee

Using identities 
\be\label{iden1}
\sinh(2\pi nbP)\sinh(2\pi n'bP)=\sum_{l=0}^{{\rm min}(n,n')-1}\sinh(2\pi bP)\sinh(2\pi b (n+n'-2l-1)P)
\ee 
and
\be\label{iden2}
{\sinh(2\pi nbP)\over \sinh(2\pi bP)}=\sum_{l=1-n,2}^{n-1}\exp(2\pi lbP)
\ee
and performing modular transformation (\ref{motrp}) we obtain:
\bea
&&Z_{(m_1,n_1),(m_2,n_2)}=\\ \nonumber
&& \sum_{l_1=0}^{{\rm min}(n_1,n_2)-1}\sum_{k_1=0}^{{\rm min}(m_1,m_2)-1}
\sum_{l=1-(n_1+n_2-2l_1-1),2}^{(n_1+n_2-2l_1-1)-1}\sum_{k=1-(m_1+m_2-2k_1-1),2}^{(m_1+m_2-2k_1-1)-1}\chi_{s+i(k/b+lb)/2}(q^{1/2}) 
\eea
in agreement with (\ref{funcf}).
Again using (\ref{giden}) and (\ref{ishid}) for
the partition function between permutation brane labeled by $(m,n)$
and product of two ZZ branes labeled by $(m_1,n_1)$ and $(m_2,n_2)$ one obtains:
\bea
&&Z_{(m,n);(m_1,n_1),(m_2,n_2)}=\int \Psi_{{\cal P}\,m,n}^{(2)}(-P)\Psi_{m_1,n_1}^{(ZZ)}(P)\Psi_{m_2,n_2}^{(ZZ)}(P)\chi_P(\tilde{q}^{2})dP=\\ \nonumber
&&\int_P {2^{3/2}\sinh(2\pi mbP)\sinh(2\pi nP/b)\sinh(2\pi m_1bP)\sinh(2\pi n_1P/b)\sinh(2\pi m_2bP)\sinh(2\pi n_2P/b)
\over (\sinh(2bP)\sinh (2P/b))^2}\\ \nonumber
&&\times\chi_P(\tilde{q}^{2})dP
\eea

Using identity  (\ref{iden1}) 
and modular transformation law for degenerate characters (\ref{motrdi})
it takes form
\bea
&&Z_{(m,n);(m_1,n_1),(m_2,n_2)}=\sum_{l_1=0}^{{\rm min}(n_1,n_2)-1}\sum_{k_1=0}^{{\rm min}(m_1,m_2)-1}\sum_{l_2=0}^{{\rm min}(n,n_1+n_2-2l_1-1)-1}
\\ \nonumber
&&\sum_{k_2=0}^{{\rm min}(m,m_1+m_2-2k_1-1)-1}
\chi_{m_1+m_2+m-2k_1-2k_2-2;n_1+n_2+n-2l_1-2l_2-2} (q^{1/2})
\eea
again in agreement with (\ref{funcf}).
This calculation can be easily generalized to the the case of generic $N$.
It shows in particularly that to produce correct formula for annulus partition function between permutation branes and products of ZZ branes the power $N$ in (\ref{permtwo}) and  (\ref{defps}) is really necessary.

\section{Defects in Liouville theory}
Defects in the Liouville theory can be constructed from the permutation branes 
$U_{{\cal P}\, s}^{(2)}(P)$ and $\Psi_{{\cal P}\,m,n}^{(2)}(P)$ on two-fold product
constructed in the
previous section using discussed in section 3 
folding trick.
 As we explained in section 3 folding trick
involves two steps, left right exchange and hermitian conjugation.
Taking into account two-point function of the Liouville theory (\ref{twopf})
one concludes that permutations branes reflection amplitude in the process
of the hermitian  conjugation should be divided by the reflection function (\ref{reflp}).
Dividing (\ref{permtwo}) and (\ref{defps}) for $N=2$ by the reflection function
(\ref{reflp}) and using (\ref{giden}) one obtains:

\be\label{defsl}
D_s(P)=U_{{\cal P}\, s}^{(2)}(P)S(-P)={\cos(4P\pi s)\over 2\sinh(2\pi bP)\sinh (2P\pi/b)}
\ee

and

\be\label{defmnl}
{\cal D}_{m,n}(P)=\Psi_{{\cal P}\,m,n}^{(2)}(P)S(-P)={\sinh(2\pi mP b^{-1})\sinh(2\pi nbP)\over \sinh(2\pi bP)\sinh (2P\pi/b)}
\ee

Now one can define
\be
X_{s}=\int_P D_{s}(P){\rm id}_{P\otimes P}dP
\ee
and
\be
X_{m,n}=\int_P{\cal D}_{m,n}(P){\rm id}_{P\otimes P}dP
\ee
where ${\rm id}_{P\otimes P}$ is the identity operator on the space $R_{{Q\over 2}+iP}\otimes R_{{Q\over 2}+iP}$.
Consider partition function with insertion of two defects parameterized  by $(m_1,n_1)$ and  
$(m_2,n_2)$
\bea
&&Z_{(m_1,n_1),(m_2,n_2)}=\int {\cal D}_{m_1,n_1}(P){\cal D}_{m_2,n_2}(P)\chi_P(\tilde{q})
\chi_P(\bar{\tilde{q}})
dP=\\ \nonumber
&&\int_P {\sinh(2\pi m_1bP)\sinh(2\pi n_1P/b)\sinh(2\pi m_2bP)\sinh(2\pi n_2P/b)
\over (\sinh(2bP)\sinh (2P/b))^2}\chi_P(\tilde{q})\chi_P(\bar{\tilde{q}})dP
\eea

Using identities (\ref{iden1}) and (\ref{iden2}) 
and performing modular transformation (\ref{motrp}) we obtain:
\bea
&&Z_{(m_1,n_1),(m_2,n_2)}=\\ \nonumber
&&\int \sum_{l_1=0}^{{\rm min}(n_1,n_2)-1}\sum_{k_1=0}^{{\rm min}(m_1,m_2)-1}
\sum_{l=1-(n_1+n_2-2l_1-1),2}^{(n_1+n_2-2l_1-1)-1}\sum_{k=1-(m_1+m_2-2k_1-1),2}^{(m_1+m_2-2k_1-1)-1}\chi_{P+i(k/b+lb)/2}(q)\chi_P(\bar{q})dP
\eea
in agreement with (\ref{deffu}).

Using identities (\ref{iden1}) and (\ref{iden2}) for fusion of defects with boundaries
and with themselves one obtains
\be
X_{m,n}|m',n'\rangle^{(ZZ)} =\sum_{l=0}^{{\rm min}(n,n')-1}
\sum_{k=0}^{{\rm min}(m,m')-1}|m+m'-2k-1,n+n'-2l-1\rangle^{(ZZ)}
\ee

\be
X_{m,n}|s\rangle^{(FZZ)} =\sum_{l=1-n,2}^{n-1}\sum_{k=1-m,2}^{m-1}|s+i(k/b+lb)/2\rangle^{(FZZ)}
\ee

\be
X_{s}|m,n\rangle^{(ZZ)}=\sum_{l=1-n,2}^{n-1}\sum_{k=1-m,2}^{m-1}|s+i(k/b+lb)/2\rangle^{(FZZ)}
\ee

\be
X_{m,n}X_{m',n'}=\sum_{l=0}^{{\rm min}(n,n')-1}\sum_{k=0}^{{\rm min}(m,m')-1}X_{m+m'-2k-1,n+n'-2l-1}
\ee
\be
X_{m,n}X_{s} =\sum_{l=1-n,2}^{n-1}\sum_{k=1-m,2}^{m-1}X_{s+i(k/b+lb)/2}
\ee

in agreement with (\ref{dbs}) and (\ref{dds}).

\section{Discussion}
We would like to outline here some directions for future work.
\begin{itemize}
\item In this paper we have constructed defects and permutation branes in the Liouville field theory,
using the classifying algebra technique.
This technique can be used to find defects and permutation branes
also in other non-rational models like $SL(2,R)$, $SL(2,R)/U(1)$, Nappi-Witten etc.

\item Another important task is to study defects and permutation branes in the Lagrangian approach
to the Liouville field theory. We can write a following mixed boundary interaction term
\be\label{mub}
 \mu_B e^{\alpha \varphi_1}e^{\beta \varphi_2}
\ee
where $\alpha(Q-\alpha)+\beta(Q-\beta)=1$,
in the product space of the two Liouville fields $\varphi_1$ and $\varphi_2$.
In the case when $\alpha=\beta$ one has the permutation symmetry.
We are tempted to think that parameter $A$, labeling continuous family in (\ref{lineq}), 
should be related to the parameter $\mu_B$ in (\ref{mub}) for this case. 
\item
The defect $X_{s_1}$ acting on  FZZ states $|s_2\rangle^{(FZZ)}$ produces the state: 
\be
X_{s_1}|s_2\rangle^{(FZZ)}=\int{\cos(4P\pi s_1)\over 2\sinh(2\pi bP)\sinh (2P\pi/b)}U_{ s_2}^{(FZZ)}(P)|P\rangle\rangle dP 
\ee
The interpretation of this state at the moment is not clear.
It would be interesting to understand this state in the matrix model approach \cite{Martinec:2003ka,Martinec:2004td}.  
\end{itemize}
\vskip4em

\noindent {\bf Acknowledgements} \\[1pt]
I am grateful to Ingo Runkel and Christoph Schweigert for useful discussions
and comments on the paper. \\
Author received partial support from the Collaborative Research Centre 
676 ``Particles, Strings and the Early Universe - the Structure of Matter and 
Space-Time''.

\end{document}